\documentstyle[11pt,moriond,epsfig]{article}

\def\Journal#1#2#3#4{{#1} {\bf #2}, #3 (#4)}

\def\be{\begin{equation}}
\def\ee{\end{equation}}
\def\bea{\begin{eqnarray}}
\def\eea{\end{eqnarray}}

\begin{document}
\vspace*{4cm}
\title{CURRENT STATUS OF COSMOLOGICAL MDM MODELS}

\author{ E.V. MIKHEEVA, V.N. LUKASH, N.A. ARKHIPOVA, A.M. MALINOVSKY }

\address{Astro Space Center of Lebedev Physical Institute,\\
 84/32 Profsoyuznaya, Moscow 117810, Russia\\}

\maketitle\abstracts{
An analysis of cosmological models in spatially flat Friedmann Universe with cosmic 
gravitational wave background and zero $\Lambda$-term is presented. The number of free 
parameters is equal to 5, they are $\sigma_8$, $n$, $\Omega_\nu$, $\Omega_b$, and $h$. 
The normalization of the spectrum of density perturbations on galaxy cluster abundance 
($\sigma_8 = 0.52\pm 0.04$) has been used to calculate numerically the value of the large 
scale CMB anisotropy ($\ell\simeq 10$) and the relative contribution of cosmological 
gravitational waves T/S. Increasing $\Omega_\nu$ weaken the requirements to the value of 
T/S, however even for $\Omega_\nu\le 0.4$ the models with $h+n\ge 1.5$ suggest considerable
abundance of gravitational waves: T/S${}^>_\sim 0.3$. In models with $\Omega_\nu\le 0.4$ 
and scale-invariant spectrum of density perturbations ($n=1$): T/S${}^>_\sim 10(h-0.47)$. 
Minimization of the value T/S is possible only in the range of the red spectra ($n<1$) and 
small $h$ ($<0.6$). It is shown that the models with T/S$\in [0, 3]$ admit both moderate 
red and blue spectra of density perturbations, $n\in[0.9,1.2]$, with rather high abundance 
hot dark matter, $\Omega_\nu\in [0.2,0.4]$. Any condition, $n<0.9$ or $\Omega_\nu<0.2$, 
decreases the relative amplitude of the first acoustic peak for more than 30$\%$ in 
comparison with its hight in the standard CDM normalized by COBE data.}

\section{Model description}

We considered a family of models with the following free parameters:
\begin{itemize}
\item $\sigma_{8}\in [0.47,0.61]$, (15 models with step 0.01);
\item $n\in [0.8,1.4]$, (7 models with step 0.1); 
\item $\Omega_\nu \in [0,0.4]$, (5 models with step 0.1);
\item $\Omega_b \in [0.01,0.11]$, (6 models with step 0.02);
\item $h \in [0.45,0.7]$, (6 models with step 0.05).
\end{itemize}

We use the analytic approximation of the transfer function by Novosyadlyj 
et al (1998)\cite{novos}. Altogether we have 18900 variants of the model. 
The derived parameters are abundance of the cold matter, $\Omega_{cm}=
1-\Omega_\nu -\Omega_b$, and the contribution of tensor mode to large-scale 
CMBR anisotropy, T/S. Our goal is to constrain the model parameters by data 
of the mass function of galaxy clusters and $\Delta T/T$ anisotropy in both 
large ($\ell\sim 10$) and small ($\ell\sim 200$) angular scales.

\section{Mass function of galaxy clusters}

The number of massive halos with the mass larger then $M$ is calculated with 
help of Press-Schechter formalism\cite{ps}. Observational data are taken from 
Bahcall \& Cen paper \cite{bcen}. 

The $\chi^2$ analysis allows us to delimit the amplitude of the power spectrum
at cluster scale with high accuracy, $\sigma_8 = 0.52 \pm 0.01$; taking into 
account the uncertainties of the Press-Schechter approximation and 
experimental systematics enhances the total errorbar by a value of $0.04$ 
\cite{eke}, \cite{borgani}. Other parameters ($n, \Omega_\nu, \Omega_b , h$)
are not constrained within their ranges by the cluster data. 

\section{CMBR Anisotropy}

The contribution of cosmic gravitational waves into the large-scale CMBR 
anisortopy is estimated by the T/S parameter:
$$
\left\langle\left(\frac{\Delta T}{T}\right)^2\right\rangle_{10^o}=\rm S+\rm T=
S\left( 1+\frac{\rm T}{\rm S}\right)\simeq 1.1\times 10^{-10}, 
$$ 
where S is the contribution of the perturbations of matter density normalized
by $\sigma_8=0.52$:
$$
\rm S=\sum_{\ell=2}^{\infty}S_{\ell}W_{\ell},\;\;\;\;
S_{\ell}=\frac{2\ell +1}{64\pi}AH_0^{n+3}
\frac{\Gamma(3-n)\Gamma(\ell+(n-1)/2)}{\Gamma^2(2-n/2)\Gamma(\ell+(5-n)/2)},\;
\;\;\;
W_{\ell}=\exp\left[-\left(\frac{2\ell+1}{27}\right)^2\right],
$$
$A$ and $W_{\ell}$ are the normalization constant and DMR window function,
respectively. The accuracy of this approximation is better than $3\%$, the
harmonics with $\ell{}_\sim^< 10$ ensure the dominant contribution.
The result of calculation of T/S$\in [0,3]$ is presented in Fig.1.
\begin{figure}
\centerline{\psfig{figure=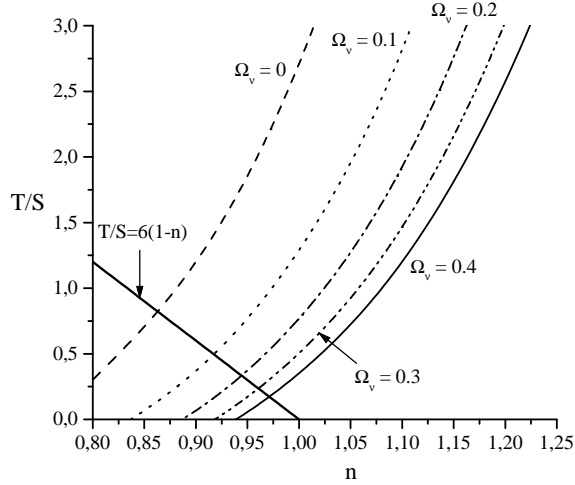,height=2.5in}}
\caption{T/S as a function of $n$ and $\Omega_\nu$ ($\sigma_8=0.52$, 
$\Omega_b=0.05$, $h=0.5$).}
\end{figure}

The value T/S icreases linearly with $h$ and decreases with $\Omega_\nu$ 
growing, therefore the curves T/S with the maximum parameter $\Omega_\nu =
0.4$ can be used to put the lower limit on T/S (see Conclusions).

Taking a moderate T/S$<0.5$ and nearly flat power spectrum ($0.92\le n\le 
1.02$), we put an upper limit on the Hubble constant, $h<0.6$, and lower limit
on the hot dark matter abundance, $\Omega_\nu >0.1$. However, the hardest 
constraint for the parameter $\Omega_\nu$ can be got when we confront 
the amplitude of the first acoustic peak in $\Delta T/T$ ($\ell\simeq 200$) 
with the observational data.    

We compare the hight of the acoustic peak generated in our models with its 
value in the standard CDM (without gravitational waves) normalized by 
the COBE data. The parameter for such a comparison is the relative amplitude 
of the peak, $\Re\equiv\Re_{\ell=200}/1.1\times 10^{-10}$, where
$\Re_\ell\equiv\ell (\ell+1)S_\ell /(\ell+0.5)$; $\Re=5.1$ for sCDM.

Evidently, $\Re$ decreases with T/S growing (and other parameters fixed).
E.g. for CDM models ($\Omega_\nu=0$ and 'standard' values for $n, \Omega_b$ and 
$h$) the relative amplitude of the acoustic peak decays by a factor 
T/S$+1\simeq 4$.

The more efficient ways to enhance the acoustic peak is a transition to the
'red' power spectra and/or high abundance of the hot matter. The role of
the 'blue' spectra becomes important when the parameter $\Omega_\nu$ rises
up (since the 'red' spectra will violate the condition T/S$\ge 0$).
The results of the derivation are presented in Fig.2. The condition
for a 'considerable' acoustic peak ($\Re\ge 4$) with the standard BBN
constraint for the baryonic density, leaves us with a broad set of
the power spectra ($n\in [0.9,1.2]$) but requires high fraction of the
hot matter ($\Omega_\nu \in [0.2,0.4]$) in the class of the models considered.
  
\section{Conclusions}

\begin{itemize}
\item The data on the galaxy cluster abundance determine the value $\sigma_8$ 
with a high accuracy, the other parameters ($n, \Omega_\nu, \Omega_b, h$) 
remain free within their ranges.    
\item None of the MDM models with $n=1$ and T/S$=0$ satisfies both 
normalizations, on the galaxy cluster abundance and large-scale $\Delta T/T$
anisotropy, which leads either to rejection from the flat spectrum or to
the introduction of a non-zero T/S (or both).
\item Small values of T/S are realised for the red spectra ($n<1$) and 
moderate $h(<0.6)$, the violation of these conditions leads to a high 
T/S(${}^>_\sim 1$).
\item 
Increasing $\Omega_\nu$ weaken the requirement to the value of T/S, however 
even for $\Omega_\nu\le 0.4$ the models with $h+n\ge 1.5$ suggest considerable
abundance of gravitational waves: T/S${}^>_\sim 0.3$.
\item 
In models with $\Omega_\nu\le 0.4$ and scale-invariant spectrum of density 
perturbations ($n=1$): T/S${}^>_\sim 10(h-0.47)$.
\item 
In models with $\Omega_b=0.05$ and $h=0.5$ we have the following approximation
for the primordial gravitational waves (the accuracy is better than 11 \% for 
$0.1\le$T/S$\le 3$):
$$
\frac{\rm T}{\rm S}=\frac{30(n-0.7)^2}{10\Omega_\nu+1}+10\Omega_\nu\left(
n^{3/2}-1.06\right).
$$ 
\item 
In double-normalized models with T/S$>$0 the hight of the acoustic peak is 
less than its `standard' value ($\Re=5.1$). The deacrease of the parameter 
$\Re$ does not exceed 30\% in models with large $\Omega_\nu \in [0.2,0.4]$ 
and any spectrum slope, $n\in [0.9,1.2]$. Any condition, $n<0.9$ or 
$\Omega_\nu<0.2$, decreases the relative amplitude of the first 
acoustic peak for more than 30$\%$ (i.e. $\Re<3.5$ in models with
$\Omega_b=0.05$, $h=0.5$). The acoustic peak practically disappears
in CDM models.
\item
When increasing the baryonic abundance the difference between $\Re$ in our models
and that in sCDM decreases. The amplitude of the acoustic peak coincides with 
its 'standard' value ($\Re\ge 4.5$) in models with 
$\Omega_b=0.1$ and either 'blue' spectrum $n\in[1,1.2]$ and 
$\Omega_\nu\ge 0.3$, 
or moderate 'red' spectrum $n\in[0.9,1]$ and $\Omega_\nu\ge 0.2$.
\item
Thus, rising the parameter $\Omega_\nu$ up to the values in the interval 
$[0.2,0.4]$ solves the problem of the first acoustic peak in $\Delta T/T$, 
leaving the baryonic density within the primordial nucleosynthesis 
constraints.   
\end{itemize}
\begin{figure}
\centerline{\psfig{figure=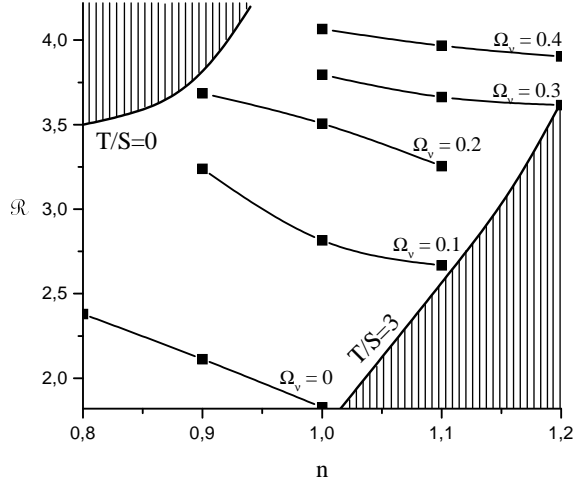,height=2.5in}}
\caption{$\Re$ as a function $n$ for different values of $\Omega_\nu$ 
($\Omega_b=0.05$, $h=0.5$, $\sigma_8=0.52$). Models in the 
non-shaded region have T/S$\in[0,3]$.}
\end{figure}

\section*{Acknowledgments}
The work was partly supported by Swiss National Foundation 
(SNSF 7IP 050163.96/1), INTAS grant (97-1192), and Russian 
Foundation ``Development and Support of Radioastronomy 
Scientific and Educational Center'' (N 315). E.V.M., V.N.L. 
and N.A.A. are grateful to the Organizing Commeittee for the 
hospitality.

\end{document}